# LSM-DFN Modeling for Seismic Responses in Complex Fractured Media: Comparison of Static and Dynamic Elastic Moduli

Ning Liu[a,*], Li-Yun Fu[b,**]

[a]*College of Mechanical and Electrical Engineering, Beijing University of Chemical Technology, Beijing 100029, China*
[b]*Key Laboratory of Deep Oil and Gas, China University of Petroleum (East China), Qingdao 266580, Shandong, China*

**ABSTRACT**

Crack microgeometries pose a paramount influence on effective elastic characteristics and sonic responses. Geophysical exploration based on seismic methods are widely used to assess and understand the presence of fractures. Numerical simulation as a promising way for this issue, still faces some challenges. With the rapid development of computers and computational techniques, discrete-based numerical approaches with desirable properties have been increasingly developed, but have not yet extensively applied to seismic response simulation for complex fractured media. For this purpose, we apply the coupled LSM-DFN model (Liu and Fu, 2020b) to examining the validity in emulating elastic wave propagation and scattering in naturally-fractured media. By comparing to the theoretical values, the implement of the schema is validated with input parameters optimization. Moreover, dynamic elastic moduli from seismic responses are calculated and compared with static ones from quasi-static loading of uniaxial compression tests. Numerical results are consistent with the tendency of theoretical predictions and available experimental data. It shows the potential for reproducing the seismic responses in complex fractured media and quantitatively investigating the correlations and differences between static and dynamic elastic moduli.

## 1. Introduction

Natural fractured reservoirs have long been an important target for the oil and gas industry. Fracture networks ranging from the microscopic to regional scales provide highly permeable conduits for fluid flow and pose a paramount influence on the mechanical and transport properties of rocks. Therefore, understanding the distribution and development characteristics of fractures at different scales is the primary factor determining the success or failure of fractured reservoir exploration and development. To determine the presence of fractures in the subsurface, geophysical exploration with seismic methods or sonic logs measurements are widely used, based on the sensitivity of wave velocities, amplitudes and spectral characteristics to fracture compliance (Vlastos, Liu, Main and Narteau, 2007). While, due to the difficulty that the distribution of fractures is complex with a wide range of spatial scales, and the density is highly variable in space, the geometrical and statistical models of fracture patterns are still debated. Many theoretical models have been proposed to analytically investigate the velocity anisotropy (including shear wave splitting) and scattering attenuation caused by constituent fractures (e.g., Crampin, 1978, 1984; Hudson, 1981, 1986), but limited to the idealization and oversimplification of complex fractures with long wave approximation for low crack densities (Vlastos et al., 2007). Nowadays, numerical method as an efficient alternative for general applicability are increasingly applied to emulate seismic responses in complex fractured media.

Numerical simulations as an efficient supplement to experimental measurements, provide an independent verification of theoretical predictions. Generally, numerical approaches are divided into two categories: Continuum methods and discrete-based models (Suiker, Metrikine and De Borst, 2001; Liu, Fu, Tang, Kong and Xu, 2020b). Conventional continuum mechanics takes matter to be continuously distributed throughout a body. It provides a reasonable assumption for analyzing the macroscale behavior of rocks by homogenization of microstructural effects. The latter regards rocks as an assembly of microstructural elements that interact with each other by microstructural forces. For fractured medium modeling, a numerical scheme with discrete fracture networks (DFNs) is increasingly used instead of the equivalent medium theories. Early continuum-based methods, e.g., the finite element method (FEM), finite difference method (FDM), and boundary element method (BEM), model the macroscopic behavior of fractured media by implementing the corresponding macroscopic constitutive models (Swoboda, Shen and Rosas, 1998; Zhao, 2015, e.g.,). Then, Lei, Latham and Tsang (2017) combine continuum-based approaches with discrete fracture network (DFN) to model fractured rocks with only a few or plenty of fractures associated with only a small amount of displacement/rotation by introducing interface elements, or joint elements (Goodman, Taylor and Brekke, 1968; Lei et al., 2017). However, modeling the high-density and complex DFNs remains difficult, which is regarded as the intrinsic limit of continuum-based methods (Jiang, Zhao and Khalili, 2017). For more complex DFNs, discrete-based approaches seem more suitable (Jing and Hudson, 2002), especially for fractured rocks with a wide range of mineral compositions and fabric anisotropies (Liu and Fu, 2020a).

*Corresponding author
**Principal corresponding author
✉ nicolaliu@buaa.edu.cn (N. Liu); lfu@upc.edu.cn (L. Fu)
ORCID(s): 0000-0001-8692-8405 (L. Fu)

Discontinuum-based methods, like molecular dynamics (MD), lattice spring model (LSM) (Hrennikoff, 1941), and discrete element method (DEM) (Cundall, 1971), have been widely used to emulate mechanical deformations in rocks (Zhao, 2010; Liu and Fu, 2020a). Inhomogeneous effects at the microlevel could be captured by these discrete-based approaches (e.g., Suiker et al., 2001; Liu et al., 2020b), where granular textures, particle-scale kinematics, and force transmission are correlated. Harthong, Scholtès and Donzé (2012) propose a coupled DEM-DFN model for strength characterization of rock masses. Then, Bonilla-Sierra, Scholtes, Donzé and Elmouttie (2015) analyze rock slope stability using DEM-DFN modeling. Based on this modeling approach, Wang and Cai (2019) emulate excavation responses in jointed rock masses. To the best knowledge of the authors, those works mainly focus on the mechanical properties and crack opening, propagation, or coalescence under static, or dynamic loadings. However, only a few works apply the discrete-based methods to emulate the wave propagation, (e.g., Suiker et al., 2001; O'Brien and Bean, 2004; Zhu, 2017; Liu et al., 2020b; Cheng, Luding, Saitoh and Magnanimo, 2020), not to mention the issues related to complex fractured media.

The aim of this paper is to attempt to combine a discrete-based numerical method with DFNs for the simulation of the elastic wave propagation and scattering in complex fractured media. Among these discrete-based methods, LSM attracts the most interest because it is flexible to model both continuous and discontinuous systems in a discrete way (Ostoja-Starzewski, 2002), and can avoid singularity-related issues in continuum-based numerical simulation methods (Pan, Ma, Wang and Chen, 2018). Moreover, unlike the DEM where elements interact through contact surfaces, the LSM connects elements by springs or beams, with several desirable properties, such as broad applicability, easy implementation, and high flexibility to handle the contact complexity of granular materials (Liu et al., 2020b). Recently, more advanced LSMs are developed to avoid the Poisson's ratio limitation of the early LSMs (Hrennikoff, 1941), e.g., Born springs (Hassold and Srolovitz, 1989), multi-body shear springs (Monette and Anderson, 1994), noncentral shear-type springs (Griffiths and Mustoe, 2001), beam element model (Karihaloo, Shao and Xiao, 2003; Lilliu and van Mier, 2003), distinct lattice spring model (DLSM) (Zhao, Fang and Zhao, 2011), and modified LSM (Liu et al., 2020b). Based on the modified LSM (Liu et al., 2020b), Liu and Fu (2020b) develop a coupled LSM-DFN model to investigate the stress-orientation effect on the effective elastic anisotropy of complex fractured media.

In this study, we introduce the coupled LSM-DFN model (Liu and Fu, 2020b; Liu, 2020) to examining the validity in emulating elastic wave propagation and scattering in naturally-fractured media. By comparison to the theoretical predictions, the implement of the modified LSM for seismic responses are validated with input parameters optimization. Base on this numerical approach, the dynamic elastic moduli are calculated and compared with static ones from quasi-static loading of uniaxial compression tests. The remaining parts of this paper are organized as follows: The methods used are introduced briefly, including the discrete-based numerical approach for seismic responses in complex fractured media, an integrated workflow of LSM-DFN modeling, and the methodology for estimating static and dynamic elastic moduli from numerical experiments in Sect. 2; to enable high-accurate modeling, the implementation of the modified LSM for seismic responses are validated with "optimal" input parameters by calibration process in Sect. 3; Section 4 presents the numerical results of LSM-DFN models with a single crack, uniformly oriented cracks, and natural fracture networks, and comparisons are made between the static and dynamic elastic moduli for the fractured samples; The conclusions and future work are underlined in Sect. 5.

## 2. Methodology
### 2.1. Numerical method: Integrated LSM-DFN scheme

To build an LSM-DFN coupling model for fractured media, homogenous matrix and fracture geometries are represented by lattice spring models (LSMs) and discrete fracture networks (DFNs), respectively. Lattice spring modeling (LSM) methods can be traced back to the 1940s, proposed by Hrennikoff (1941) to solve the elasticity problem. Then, the lattice model is introduced to calculate the effective mechanical properties (Nayfeh and Hefzy, 1978), distributed disorder influence (Curtin and Scher, 1990a), stress concentrations and toughness increases (Curtin and Scher, 1990b). With the rapid development of computers and computational techniques, LSM is applied to the study for other mechanical and dynamic behaviors, e.g., tectonic processes (Saltzer and Pollard, 1992; Mora and Place, 1998), and fracture problems (Schnaid, Spinelli, Iturrioz and Rocha, 2004; Kosteski, D'ambra and Iturrioz, 2009). To overcome the fixed Poisson's ratio limitation existing in conventional LSMs, some advanced models are developed. For example, Liu et al. (2020b) modify the LSM by introducing an independent micro-rotational inertia to avoid the Poisson's ratio limitation and improve the LSM in characterizing scale-dependent effects. Then, Liu and Fu (2020a) use this modified LSM to study the elastic characteristics of digital cores, to incorporate with random void model (RVM) for organic-rich shale modeling (Liu, Fu, Cao and Liu, 2020a), develop a coupled LSM-DFN model to investigate the stress-orientation effect on the effective elastic anisotropy of complex fractures (Liu and Fu, 2020b) and to consider rough contact deformation for fracture surfaces (Liu, 2020). While, all the works remain quasi-static elastic issues.

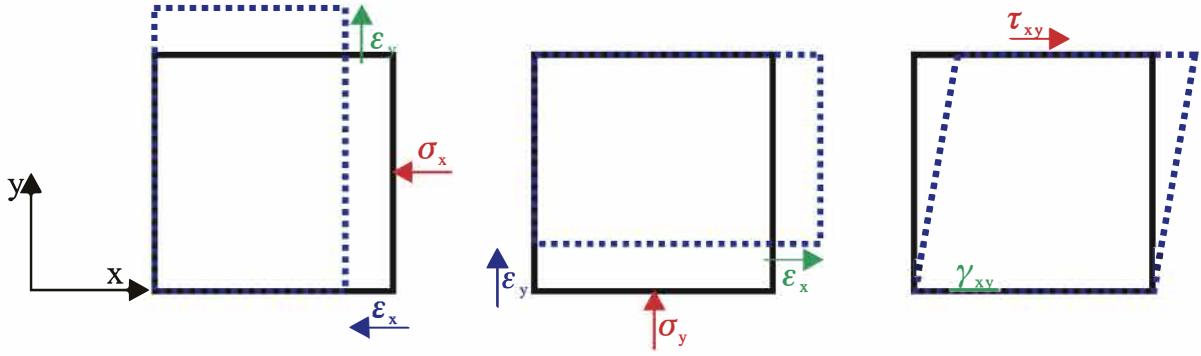

**Figure 1**: Schematic definition of the elastic constants in two dimensions.

Here, we introduce the LSM-DFN modeling to emulate elastic wave propagation and scattering in naturally-fractured media. Firstly, we emulate the matrix as a linearly elastic, homogeneous, and isotropic material using the regular triangular modified LSM. For a digital image of fractured rocks, the lattice nodes are set in an approximate zone represented by digital pixels. DFNs can be stochastically created (Hyman, Gable, Painter and Makedonska, 2014), or extracted from a digital image (Liu and Fu, 2020b; Liu, 2020). The resulting DFNs are incorporated into the LSMs to form the LSM-DFN model for fractured rocks. According to stress-free assumption for crack surfaces (Murai, 2007), the corresponding nodal interactions crossing the DNFs are removed from the model (Lisjak and Grasselli, 2014; Liu and Fu, 2020b).

### 2.2. Estimation of static elastic modulus: Uniaxial compression test

The determination of static elastic modulus is based on the measurement of the relation between elastic deformation and the known force (Ciccotti and Mulargia, 2004). For a two-dimensional (2D) case, three independent constants characterize the elasticity of the transversely isotropic model in principal coordinates, as schematically shown in Figure 1. the constitutive relations can be given by,

$$\begin{pmatrix} \varepsilon_x \\ \varepsilon_y \\ \gamma_{xy} \end{pmatrix} = \begin{pmatrix} \frac{1}{E_x} & -\frac{\nu_{yx}}{E_y} & 0 \\ -\frac{\nu_{xy}}{E_x} & \frac{1}{E_y} & 0 \\ 0 & 0 & \frac{1}{2\mu_{xy}} \end{pmatrix} \begin{pmatrix} \sigma_x \\ \sigma_y \\ \tau_{xy} \end{pmatrix}. \qquad (1)$$

Here, we conduct uniaxial compression tests to calculate the stress-strain responses of specimens and obtain the Young's moduli and the pair of Poisson's ratios in Eq. (1) by the slopes of the response curves,

$$E_x = \frac{\sigma_y}{\varepsilon_x}, \qquad (2)$$

$$E_y = \frac{\sigma_y}{\varepsilon_x}; \qquad (3)$$

$$\nu_{xy} = -\frac{\varepsilon_y}{\varepsilon_x}, \tag{4}$$

$$\nu_{yx} = -\frac{\varepsilon_x}{\varepsilon_y}; \tag{5}$$

and

$$E_x \nu_{yx} = E_y \nu_{xy}, \tag{6}$$

owing to the known symmetry condition. It is worth noting that Saint Venant (1863) provides an approximation of the shear moduli in orthotropic materials, based on the values of the rest of the elastic constants,

$$\frac{1}{\mu_{ij}} = \frac{1}{E_i} + \frac{1 + 2\nu_{ji}}{E_j} \quad (i, j = x, y). \tag{7}$$

### 2.3. Estimation of dynamic modulus: Seismic responses

Dynamic elastic moduli can be derived from elastic-wave velocity and rock density. Based on the seismic responses, we can measure velocity of a stress wave passing through a material. For carrying out this seismic pulsing method, a vertical downward displacement excitation, Ricker wavelet, given by

$$u_z = -A_0 \left(1 - 2\pi^2 f^2 (t-T)^2\right) \exp\left(-\pi^2 f^2 (t-T)^2\right), \tag{8}$$

where $A_0$, $T$ and $f$ represent amplitude, period, and central frequency, respectively, is applied at the center of the upper surface, and the receivers are arranged as shown in Figure 2. The velocities of the compressional and shear propagation of elastic waves covering the certain distance for a known receiver can be calculated by the time between the sending and receiving waves. According to the classical theory of elastic wave propagation, the compressional and the shear wave velocity, $c_P$ and $c_S$ are given by,

$$c_P = \sqrt{\frac{\lambda + 2\mu}{\rho}}, \tag{9}$$

$$c_S = \sqrt{\frac{\mu}{\rho}}, \tag{10}$$

where $\rho$ is the density. The Young's modulus in 2D can be obtained by

$$E = \frac{4\mu(\lambda + \mu)}{\lambda + 2\mu}, \tag{11}$$

so we could estimate the dynamic Young's modulus by,

$$E_{\text{dynamic}} = (c_S)^2 \left(2 - \left(\frac{c_S}{c_P}\right)^2\right) \rho. \tag{12}$$

## 3. Validation and Calibration

The objective of this section is to validate the implement of the modified LSM for seismic responses. By calibration process, this LSM-DFN model could enable high-accurate modeling by properly choosing reasonable input parameters, e.g., meshing resolutions, stability conditions, and numerical damping (Liu and Fu, 2020b). Poisson's ratio $\mu$ is a key index for two-dimensional parameter identification and directly related to the ratios of the shear to normal stiffness $\xi = \frac{K_s}{K_n}$ and the compressional to shear velocity $\frac{c_P}{c_S}$, by

$$\nu = \frac{K_n - K_s}{3K_n + K_s} = \frac{1-\xi}{3+\xi}, \tag{13}$$

and

$$\left(\frac{c_P}{c_S}\right)^2 = \frac{3K_n + K_s}{K_n + K_s} = \frac{3+\xi}{1+\xi}. \tag{14}$$

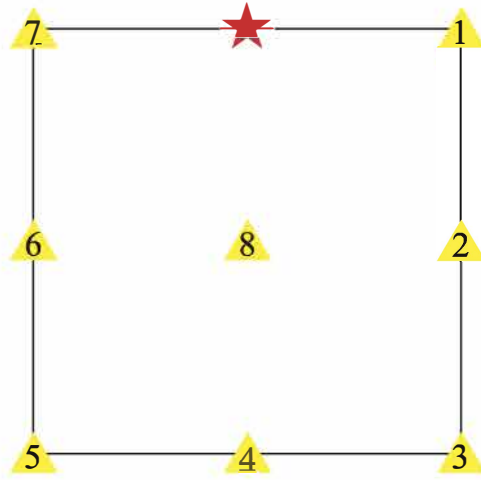

Figure 2: Source-receiver placement of seismic response simulation: The source and the receivers are marked by red star and yellow triangles.

**Table 1**
Material properties in the simulation

| Parameters | Density $\rho$ (kg/m³) | Elastic constants | | Theoretical wave velocities | |
|---|---|---|---|---|---|
| | | $E$ (GPa) | $\mu$ | $v_P$ (m/s) | $v_S$ (m/s) |
| Value | 2500 | 70 | 0.33 | 5612.49 | 3240.37 |

More details are given in our previous works (Liu et al., 2020b). We perform a comprehensive series of numerical simulations to explore the "optimal" inputs for the subsequent studies. The physical and mechanical properties of lattice nodes are reported in the Table 1.

### 3.1. Lattice spacing and meshing sensitivity

Lattice arrangement of discrete-based numerical methods may cause numerical anisotropy. Liu and Fu (2020b) study the anisotropic effect and meshing sensitivity of the LSMs by Brazilian tests. With a higher resolution, the anisotropy caused by lattice orientation could be minimized, but with heavy computational costs. In that work, Liu and Fu (2020b) give a suggestion for meshing resolution of a round specimen with a diameter of $D$ and lattice spacing of $d$: The maximum errors of models are 1.09% for $D/d = 400$, and 0.22% for $D/d = 600$. Therefore, we build models with a resolution of $D/d = 600$ for more accurate simulations.

### 3.2. Time step and source frequency

The central difference method is conditionally stable. To ensure the stability of this explicit integration algorithm applied in the modified LSM, the critical time step $\Delta t_{cr}$ is employed by

$$\Delta t \leq \Delta t_{cr}, \tag{15}$$

LSM-DFN modeling for seismic responses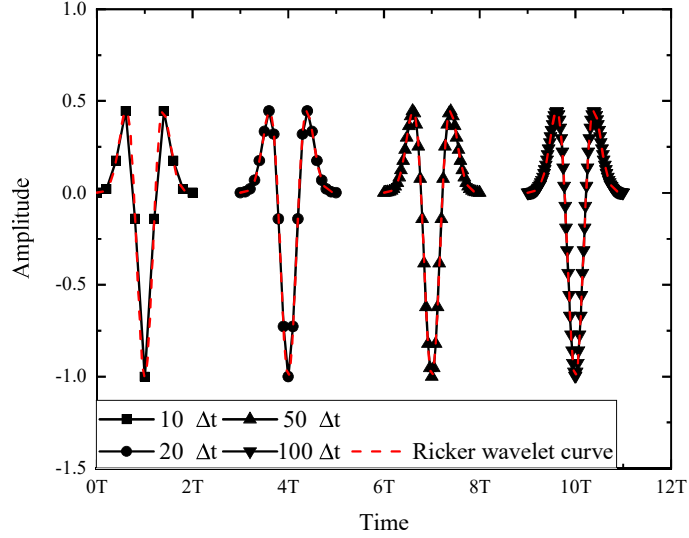

**Figure 3**: Time steps and excitation signal of Ricker wavelet.

$$\Delta t_{\text{cr}} = \frac{2}{\omega_{\max}} = \frac{T_{\min}}{\pi}, \tag{16}$$

where $\omega_{\max}$ and $T_{\min}$ are the largest eigen-frequency and smallest natural period within the lattice spring networks. Due to the fact that the smallest natural period $T_{\min}$ is always larger than or equal to the smallest natural period of the smallest-size element $\min(T^{(e)})$, we could estimate $T_{\min}$ by

$$\Delta t_{\text{cr}} \approx \frac{\min(T^{(e)})}{\pi} = \frac{\min(\frac{d}{C})}{\pi}, \tag{17}$$

where $d$ is lattice spacing and sonic speed $C$ by

$$C = \sqrt{\frac{E}{\rho}}. \tag{18}$$

Therefore, in this work, the critical time step $\Delta t_{\text{cr}}$ is defined by

$$\Delta t_{\text{cr}} \approx \frac{L}{600\pi}\sqrt{\frac{\rho}{E}} \approx L \times 10^{-7}, \tag{19}$$

in which $L$ is the width/length of the square sample. Too small time step increases the calculation time, whereas too large time step cannot satisfy the stability condition, or cannot provide enough data to form an excitation signal of Eq.(8) for a given frequency, either. Therefore, the time step limits the center frequency range for seismic responses. Figure 3 shows the applied excitation signal with different time steps. We can get a better fitting-excitation with more time steps, generally

$$\frac{T}{\Delta t} \geq \frac{20}{2}, \tag{20}$$

namely the period in Eq. (8) is more than 10 times as long as the time step.

### 3.3. Numerical damping

For discrete-based numerical methods, like molecular dynamics (MD) simulations, and DEM developed by Cundall and Strack (1979), the dynamic relaxation scheme with an artificial damping proposed by Otter, Cassell and Hobbs

Page 6 of 18

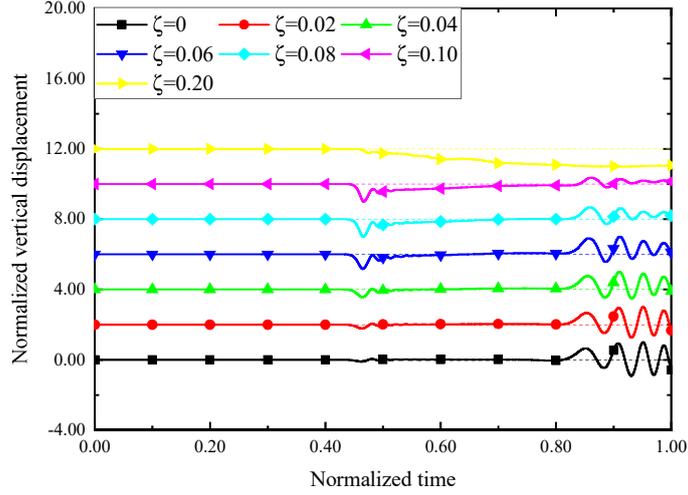

Figure 4: Vertical displacement responses at Receivers 1 and 7 under different numerical damping.

(1966) is widely used. With the numerical damping $\zeta$, the residual kinetic energy of the lattice node is dissipated for a stable numerical solution, where the dissipated force $\Delta\mathbf{G}_{\text{dissipated}}$ of node $i$ may be written as follows,

$$\frac{(\Delta\mathbf{G})^i_{\text{dissipated}}}{\mathbf{G}_i} = -\zeta\,\text{sgn}\left(\dot{\mathbf{x}}_i^{\left(t-\frac{\Delta t}{2}\right)} + \frac{\ddot{\mathbf{x}}_i \Delta t}{2}\right), \tag{21}$$

$$\mathbf{G} = \begin{pmatrix} \mathbf{F} & \mathbf{M} \end{pmatrix}^T, \tag{22}$$

$$\mathbf{x} = \begin{pmatrix} \mathbf{u} & \theta \end{pmatrix}^T, \tag{23}$$

where $\mathbf{G}$ and $\mathbf{x}$ are vectors of generalized forces and generalized displacements, respectively (Liu et al., 2020b). Figure 4 illustrates the vertical displacement spectra in time-domain for Receivers 1 and 7 shown in Figure 2. It can be seen from the figure that the displacement signal oscillates and does not stabilize in a long time when the numerical damping is 0. It is difficult to distinguish the wave types and determine the corresponding arrival times. When the damping is larger than, or equal to 0.20, the signals are attenuated significantly to almost zero within a relatively short period, and the waveforms are severely disordered. The corresponding arrival times of compressional or shear wave cannot be identified, either. From this figure, we could roughly estimate that the ideal numerical damping should be in the range of 0.04 to 0.1. The numerical damping is not only subject to time step but also lattice spacing. Here, we choose 0.06 as the input value of the numerical damping for seismic response simulation.

Summarily, in this work, the time-step and the non-viscous damping for the numerical models with a size of 600m × 600m are $\Delta t = 0.1$ $\mu$s, and 0.06, respectively. Figure 5 shows the snapshots of the wavefields at $t = 0.1$, 0.2, 0.5, 1.0 s, sent by an excitation with a period of $T = 600$ $\mu$s. The compressional and shear waves can be clearly identified from the snapshots, and we can estimate the ratio of compressional- to shear-wave velocity is approximately $\sqrt{3}$, consistent with the theoretical predictions by Eq.(14). This numerical result further verifies the implementation of the LSM and its general application to seismic response simulation.

## 4. Numerical Results and Discussions

In this section, to demonstrate that this integrated workflow of LSM-DFN can be used to emulate elastic wave propagation and scattering in fractured media, we conduct the following numerical experiments for the static and dynamic moduli of the material listed in Table 1. The wavefields and the uniaxial compression samples are in 600 m × 600 m homogeneous, isotropic, linearly elastic domain with a single crack (Case 1), uniformly oriented cracks (Case 2), and natural fracture networks (Case 3). Ricker wavelet point source given by Eq. (8) with a peak period of $T = 600$ $\mu$s and receivers are located as Figure 2.

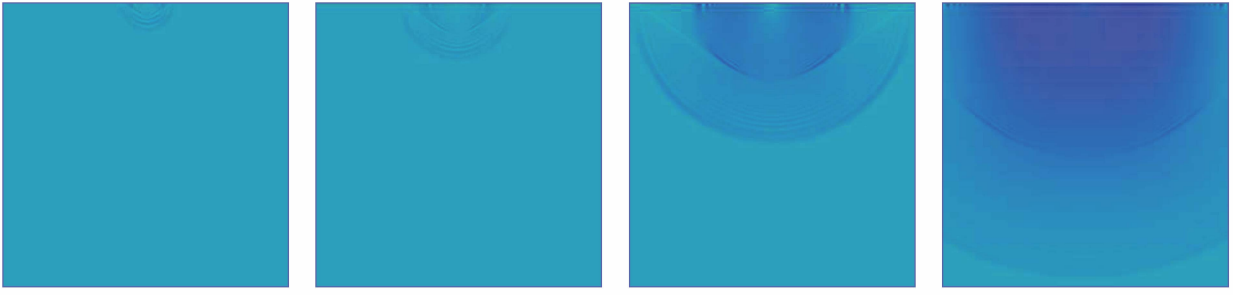

**Figure 5**: Snapshots of the vertical displacement wavefield using modified LSM at $t = 0.10, 0.20, 0.50, 1.00$ s.

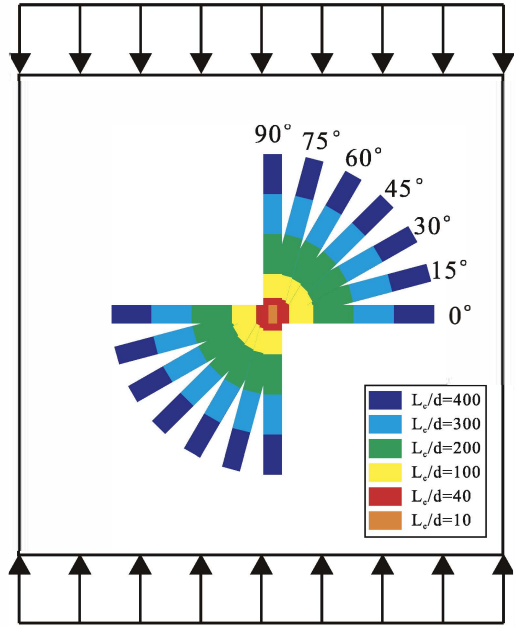

**Figure 6**: Uniaxial compression tests of LSM-DFN models with six different half-lengths of 5, 20, 50, 100, 150, and 200 m and seven different inclined angles of 0°, 15°, 30°, 45°, 60°, 75°, and 90°.

### 4.1. Case 1: LSM-DFN with a single crack

In this case, we couple a single crack located at the center of the domain into the lattice spring networks. The single crack is designed with six different half-lengths of 5, 20, 50, 100, 150, and 200 m and seven different inclined angles of 0°, 15°, 30°, 45°, 60°, 75°, and 90°. Totally, $6 \times 7 \times 2 = 84$ numerical experiments are conduct for quasi-static loadings and seismic responses. Figure 6 presents the loading patterns under uniaxial compression for static moduli.

We hold the LSM-DFN model with a half-length of 200 m up as an example shown in Figure 7: From left to right, we present the differential stress fields under quasi-static loading state at $\varepsilon_y = 0.60\%$, and snapshots of the vertical displacement wavefield at $t = 0.7$ s. From this figure, we can see static stress concentration appears at the tips of these cracks under uniaxial compression tests as theoretically predicted; a large fracture hinders the wave propagation and compressional wave are reflected upwardly at the crack interface for seismic responses.

For a two-dimensional homogeneous and isotropic material containing a single crack, the effective Young's modulus $E$ in the direction of the loading is given by Walsh (1965),

$$\frac{1}{E} = \frac{1}{E_{\text{intact}}} \left( 1 + \frac{2\pi L_c^2}{A} \cos^2 \beta \right), \tag{24}$$

where $E_{\text{intact}}$, $L_c$, $\beta$, and $A$ is the effective Young's modulus of the intact sample, the half-length of a crack, the inclined angle with the horizonal axis, and the area of a specimen, respectively. Figure 8 shows the counter plots of the effective

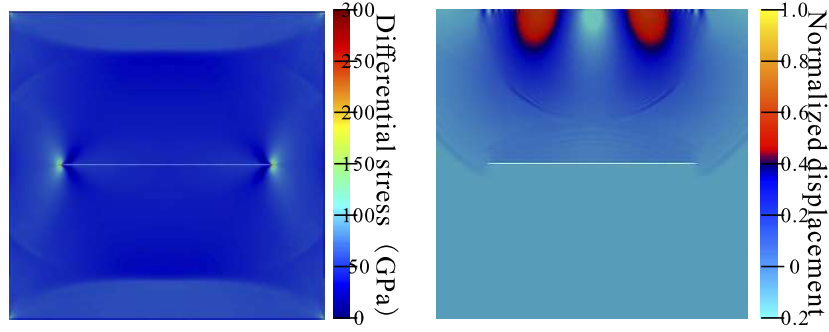

**Figure 7**: LSM-DFN model with a single crack (crack length equal to 400 m): Differential stress fields under quasi-static loading state at $\varepsilon_y = 0.60\%$, and snapshots of the vertical displacement wavefield at $t = 0.70$ s (from left to right).

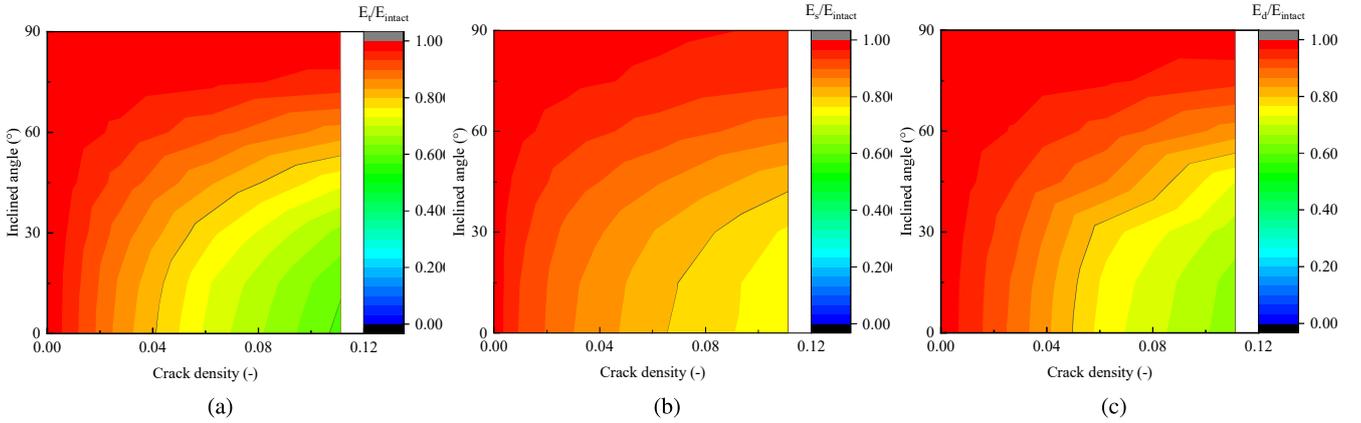

**Figure 8**: Effective Young's moduli of cracked medium containing one crack with respect to crack density and inclined angle: (a) theoretical predictions; (b) static value from uniaxial compression tests; (c) dynamic value from sonic measurements of Receiver 4.

Young's moduli of cracked medium containing one crack for theoretical predictions (Eq.(24)), and quasi-statically (Eq. (3)) and dynamically numerical simulations (Eq. (12)), where crack density $\varepsilon$ is given by Huang, Chandra, Jiang, Wei and Hu (1996)

$$\varepsilon = \frac{N L_c^2}{A}, \tag{25}$$

where $N$ is the total number of cracks, and $N = 1$ for this case. The three subfigures show similar tendency of the effective Young's moduli as a function of crack density and inclined angle, which could verify the implement of this LSM-DFN scheme for static and seismic numerical experiments. While, with increasing crack density, the static Young's moduli show higher values. That's because crack closure occurs due to a higher pressure from a larger compressive strain and shows the associated stiffening of the cracked medium. In fact, the equivalent anisotropy obtained by quasi-static uniaxial compression includes both the inherent and stress-induced parts (Liu and Fu, 2020b).

In order to better show the difference between the static and dynamic elastic moduli obtained by the numerical simulation and the theoretical prediction, Eq. (24) is rewritten as,

$$\frac{E}{E_{\text{intact}}} = \frac{1}{1 + \frac{2\pi L_c^2}{A}\cos^2\beta} = \frac{1}{1 + \frac{2\pi}{A}\left(L_c\right)_x^2}, \tag{26}$$

where $\left(L_c\right)_x$ is the value of crack half-length projected onto the horizontal axis perpendicular to compression direction. We can see the equivalent elastic modulus of the medium with a single crack is directly related to $\left(L_c\right)_x$. In Figure 9, the solid and hollow symbols labeled with suffixes of "-s" and "-d" indicate static and dynamic elastic moduli, respectively. Specifically, taking the solid and hollow squares labeled with "5-s" and "5-d" as examples, they represent the numerical data of an LSM-DFN model with a crack half-length of 5 m from uniaxial compression tests and seismic responses, respectively. Generally, the dynamic elastic moduli are in good agreement with the theoretical prediction values. The numerical data appears near the theoretical prediction curve just with slight disturbances. It is worth noting that in the zoom graph (with background color of cyan) in Figure 9 the static moduli of LSM-DFN models with crack inclined angles of 90° deviates greatly from the theoretical predictions and the corresponding dynamic results. This phenomenon may be attributed to the crack opening in transverse direction due to compression. It violates the assumption that the crack does not deform in the theoretical prediction, which leads to the equivalent stiffness values being lower than the theoretical prediction results. Therefore, for this case, the results obtained based on the seismic responses can better capture the intrinsic anisotropy of cracked media.

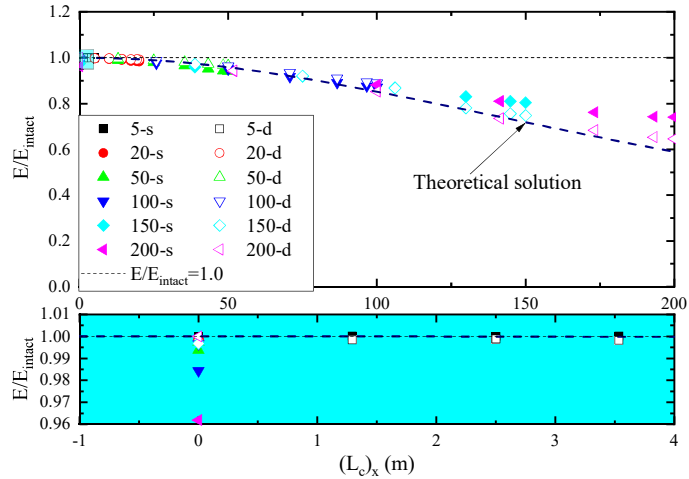

Figure 9: Normalized effective Young's modulus with respect to the value of crack half-length projected onto the horizontal axis: Solid and hollow symbols mean quasi-static and dynamic values, respectively.

For quasi-static problems, the differential stress fields obtained in Figure 10 are similar to those shown in Figure 7, with stress concentration and stress shielding appearing at the crack tips and surfaces. While, there are obvious differences between different wavefield snapshots as shown in Figure 11, indicating that geophysical exploration based on seismic responses may be better to quantify the impact of fracture characteristics on the mechanical properties.

### 4.2. Case 2: LSM-DFN with uniformly oriented cracks

In this case, we extend the single-crack numerical experiments in Case 1 to a set of four-crack LSM-DFN models in three types of spatial arrangements. The crack centers are arranged in a collinear, stacked, and mixed manner. These three spatial arrangements of four cracks with a half-length of 50 m and seven different inclined angles of 0°, 15°, 30°, 45°, 60°, 75°, and 90°. Totally, $7 \times 3 \times 2 = 42$ numerical experiments are addressed for quasi-static loadings and seismic responses. Figure 12 shows the differential stress fields under quasi-static loading state at $\varepsilon_y = 0.60\%$, and snapshots of the vertical displacement wavefield at $t = 0.70$ s of the LSM-DFN models with inclined angle of 0° in three types of spatial arrangements. It can be seen from the differential stress field distribution of the cracked media with three types of spatial arrangements, the tips of the coplanar cracks approach closest to each other. As expected, the stress is concentrated at the area between the tips of the cracks. For the media with stacked crack surfaces closest to each other, the stress shielding dominates the differential stress field distribution. The local maximum of the differential stress field of the media with mixed arrangement is between the first and second arrangements. The seismic wavefield snapshots of the three kinds of fractured media as shown in Figure 12 could further explain the phenomenon of dynamic stress concentration from the perspective of elastic wavefield theory, that is, dynamic stress concentration is the result of elastic wave diffraction, or we could call it elastic wave scattering as well. Pao and Mow (1971) ever mentioned that wave scattering has a broader implication than the original meaning of wave diffraction. When the elastic wave bypasses the crack tips, the diffracted part dominates the scattered waves. Each component of the crack can be regarded as the center of the secondary disturbance. Around the normal of the incident wave front, the intensity of the spherical wave diverging outward does not change suddenly with respect to direction. The perturbation can be taken as the sum of the perturbations caused by all the secondary waves. The phase of each secondary wave is delayed by a quantity corresponding to the distance from the center of the receiver, so the arriving secondary waves interfere with each other, resulting in diffraction (Pao and Mow, 1971).

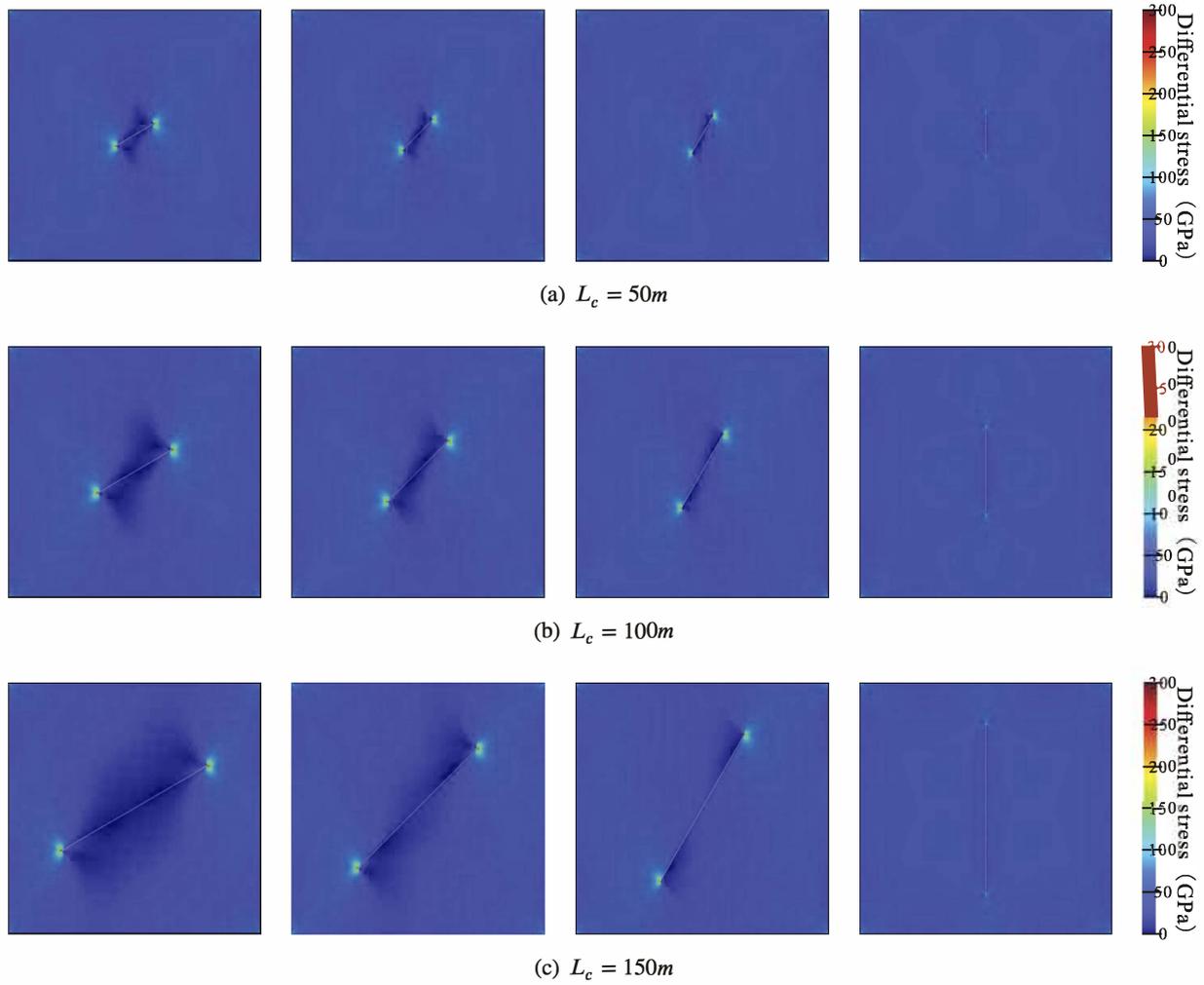

**Figure 10**: Differential stress fields under quasi-static loading state at $\varepsilon_y = 0.60\%$ of LSM-DFN models with crack half-lengths of 50 m, 100 m, 200 m and four different inclined angles of 30°, 45°, 60°, and 90°.

Figure 13 shows a comparison made between the theoretical, static, and dynamic moduli. It can be seen from the figure that the change trend of the static elastic moduli is in good agreement with the theoretical values. The models with the first and second crack arrangements can be seen based on the Voigt and Reuss bound theories, respectively. The cracked part can be regarded as a material layer with a lower stiffness. The effective moduli of the models with the third spatial arrangement are within the zone covered by Voigt-Reuss bounds as theoretical predictions. These numerical experimental results further verify the effectiveness of the LSM-DFN models. However, due to stress stiffening caused by uniaxial compression, the static Young's modulus values are slightly higher than the theoretically predicted ones. However, for Case 2, there are obvious deviations between the elastic moduli obtained from the seismic responses and the theoretical predictions, especially for the dynamic elastic modulus calculated from Receive 4. The phenomenon is related to the receiver placement for the estimation of dynamic elastic moduli. Specifically, for Case 1, Receiver 4 happens to be along the line between the excitation and crack center, and directly below the center of the crack as shown in Figure 2, so we could accurately estimate the dynamic elastic moduli. While, for the first arrangement of Case 2, along the path from the excitation to Receiver 4, seismic wave doesn't pass through cracks as shown in Figure 12(a), so the estimated dynamic elastic moduli are approximately consistent with the value of homogeneous media. For the second arrangement of Case 2, the crack centers are arranged along the path from excitation to Receiver 4 in a stacked manner. From the wavefield snapshots as shown in Figure 12(b), stacked cracks prevent Receiver 4 from receiving the direct seismic signal excited by the source, so the dynamic elastic modulus cannot be accurately estimated based on this receiver. Figure 13(b) shows the average dynamic elastic moduli obtained from 3 receivers, Receivers 3-5 where the tendency seems more reasonable. Therefore, we may conclude that the obtained dynamic elastic moduli mainly depend on the receiver arrangement and accurate travel-time estimation.

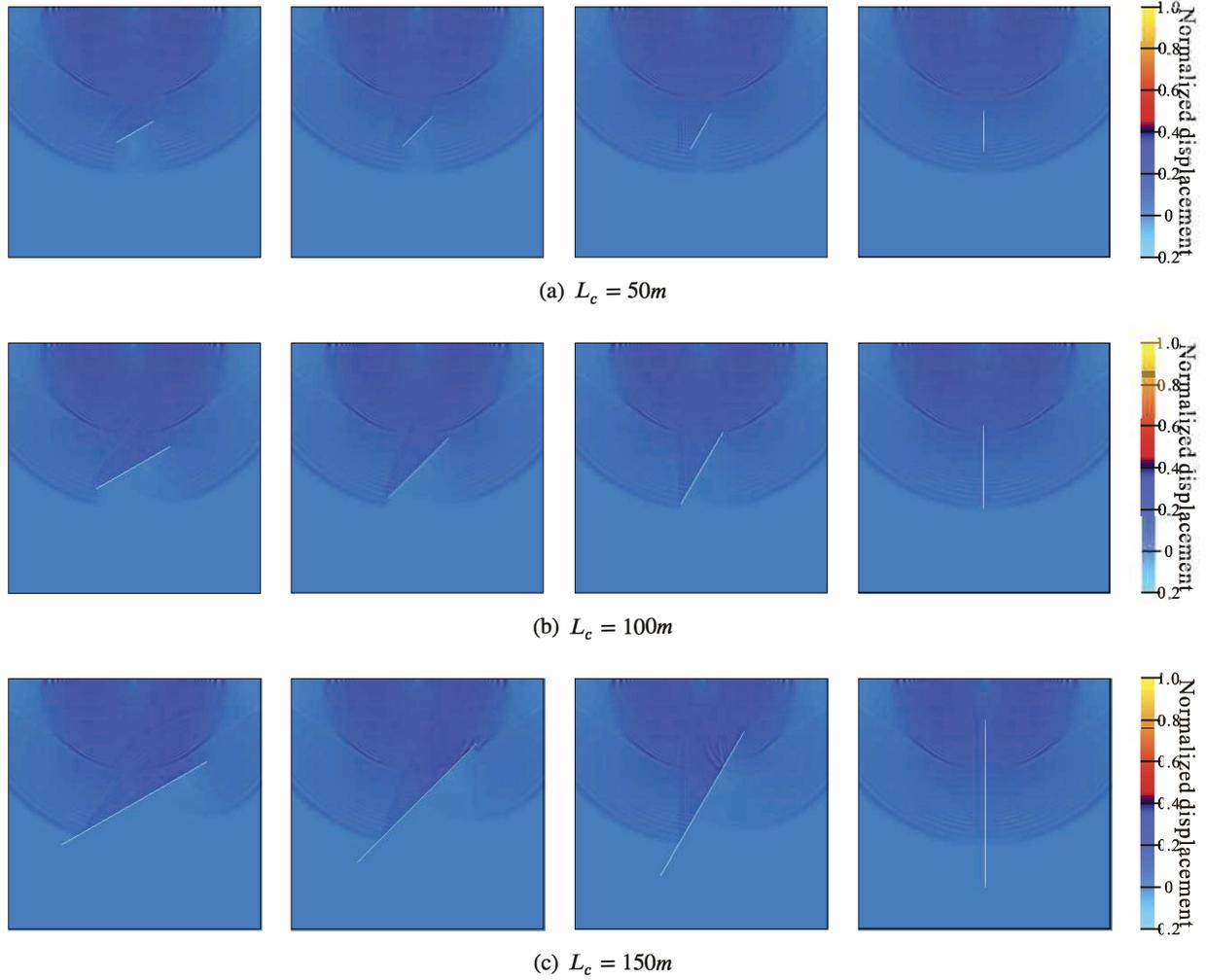

**Figure 11:** Snapshots of the vertical displacement wavefield at $t = 0.70$ s: LSM-DFN models with crack half-lengths of 50 m, 100 m, 200 m and four different inclined angles of 30°, 45°, 60°, and 90°.

Figure 14 and Figure 15 show the differential stress fields under quasi-static loading state at $\varepsilon_y = 0.60\%$, and snapshots of the vertical displacement wavefield at $t = 0.70$ s of the LSM-DFN models with inclined angles of 30°, 45°, 60°, and 90° in three types of spatial arrangements. It can be seen from Figure 14 that as the distance between the crack tips decreases, the stress concentration phenomenon increases. From the perspective of the seismic wavefield, Figure 15 further explains the diverse degrees of static and dynamic stress concentration caused by fracture arrangements and inclined angles. Different crack arrangements with various inclined angles make the medium show varied equivalent elastic modulus, resulting in uneven differential stress field distribution, and around the crack tips show distinct static and dynamic stress concentrations. Such numerical simulations provide valuable insights into the mechanisms and processes related with stress variation and seismic responses. Moreover, they could help assess theories by comparing simulations with predictions, serving as powerful tools to improve both theory and experiment.

### 4.3. Case 3: LSM-DFN modeling for naturally-fractured reservoirs

According to Case 1 and Case 2, we examine the validity in emulating elastic wave propagation and scattering in cracked media using LSM-DFN model and analyze the effects of the receiver placement on the dynamic elastic moduli. In this case, we apply the integrated workflow of LSM-DFN modeling proposed by Liu and Fu (2020b) to seismic responses in naturally-fractured reservoirs. Figure 16 shows two real outcrops of Ordovician carbonates in the northwest of the Shuntuoguole low uplift of Tarim Basin in China. The complex fractured reservoirs with different scales and orientations manifest multi-stage tectonic movements (Liu and Fu, 2020b).

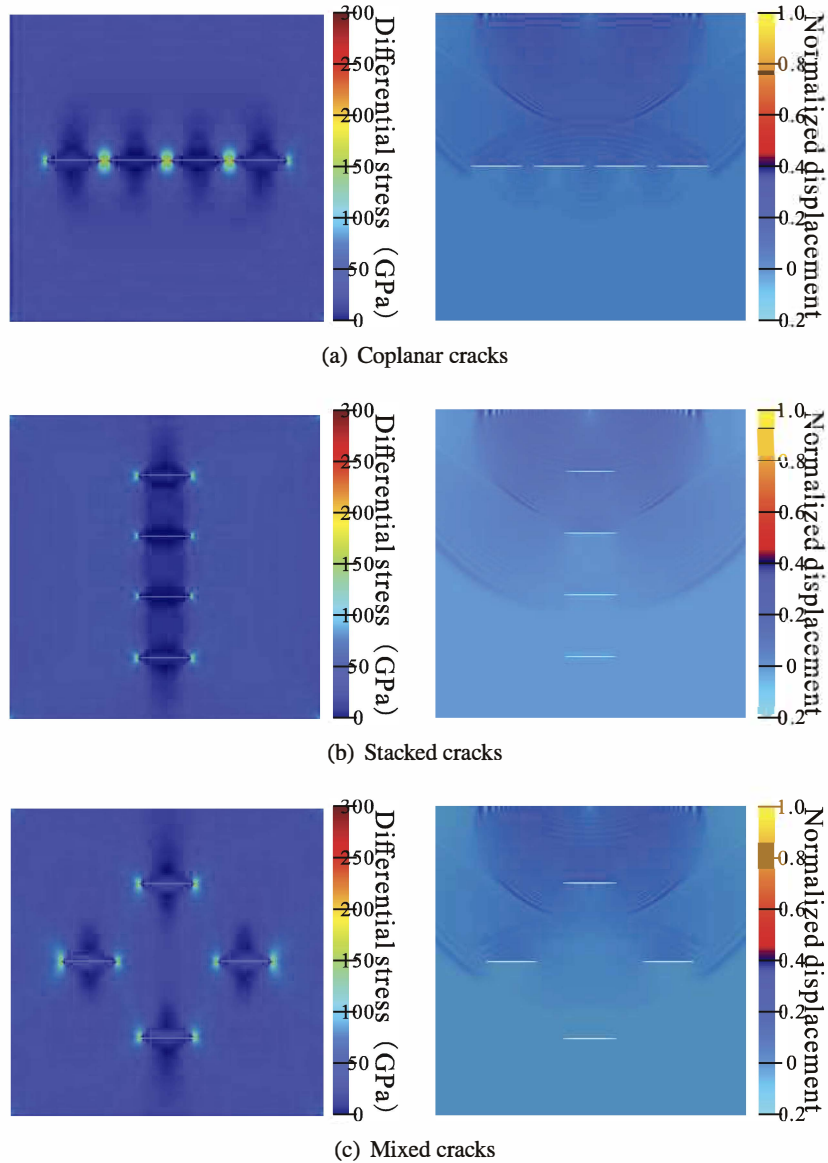

**Figure 12**: LSM-DFN model with a set of four cracks with a half-length of 50 m and inclined angle of 0° in three types of spatial arrangements, (a) coplanar, (b) stacked, and (c) mixed cracks: Differential stress fields under quasi-static loading state at $\varepsilon_y = 0.60\%$ , and snapshots of the vertical displacement wavefield at $t = 0.70$ s (from left to right).

The complex fractured systems raise challenges in high-accurate modeling for natural fracture networks. Liu and Fu (2020b) extracted manually from a digital photograph by the introduction of Healy, Rizzo, Cornwell, Farrell, Watkins, Timms, Gomez-Rivas and Smith (2017). Then, Liu (2020) improve the fracture extraction by image processing method, the gradient Hough transform (GrdHT) introduced by Mukhopadhyay and Chaudhuri (2015). Figure 17 displays the flow diagrams for efficient line detection, where those fracture networks could be separated from the image background with a gradient magnitude less than a certain threshold. Figure 17(a) shows smaller-scale cracks relative to the overall model size, whilst Figure 17(b) contains multi-scale cracks. In Figure 17(b), those large-scale cracks almost cross the whole sample and divide the sample into nearly intact large blocks. The effective stiffness of the blocky system is seen as stronger than those of samples with homogeneously distributed small fractures (Harthong et al., 2012). Figure 18 shows the distribution characteristics of fracture length of the two outcrops. We get 171 and 619 fractures from the two outcrops, respectively. The fracture lengths of Outcrop 1 are mainly concentrated in a small interval of 4 ∼ 135 m, and those of Outcrop 2 are scattered in a larger interval of 2 ∼ 380 m. The different mechanisms caused by fracture scales on stress fields and seismic responses will be illustrated in the following parts.

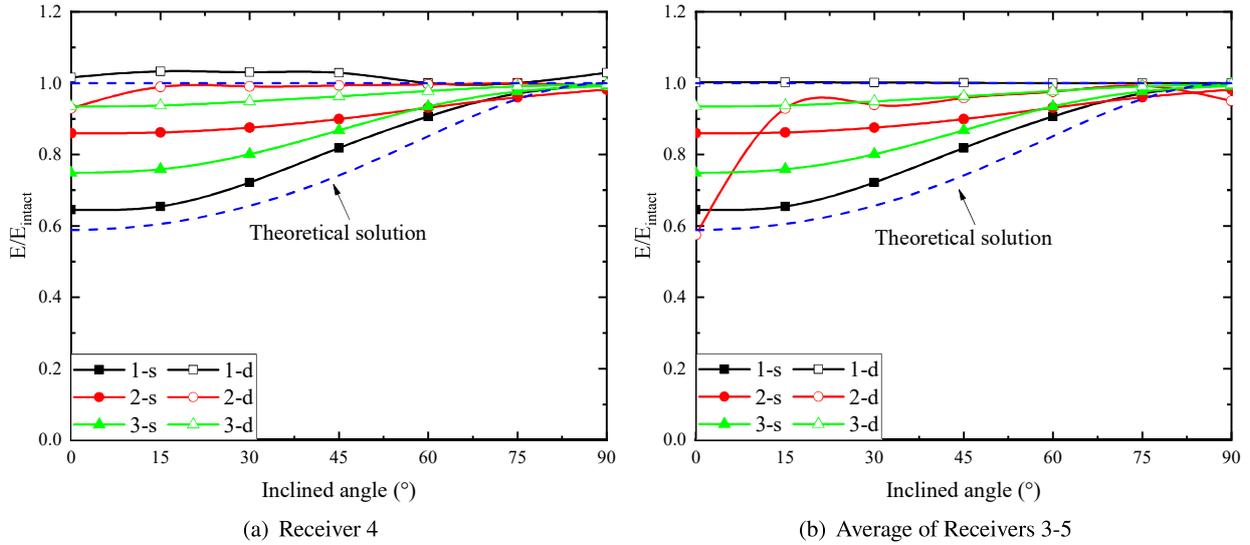

Figure 13: Normalized effective Young's moduli of cracked medium in three types of spatial arrangements as a function of inclined angle with dynamic moduli from (a) Receiver 4 and (b) Average of 3 receivers, Receivers 3-5.

Figure 19 and Figure 20 show the differential stress fields and seismic responses of Outcrop 1 and Outcrop 2. The natural fracture networks cause complex stress interaction, and strong stress concentration occurs where the tips of multiple fractures cluster. The snapshots of seismic wavefields help comprehend the dynamic stress concentration mechanism in complex fractured media from the perspective of elastic wave scattering. Figure 20(b) shows that the large-scale cracks that almost cross the entire sample hinder the propagation of seismic waves. It causes obvious wave diffractions and wave reflections within the blocky system near the excitation source. From the snapshots, we can see the seismic wave is prevented from continuing to propagate. In fact, we could attribute it to the subsequent material points don't move, namely, that part dose not deform. That can explain a stronger stiffness of the blocky system compared to the samples with homogeneously distributed small fractures. Contrastively, for Outcrop 1 with smaller-scale cracks, the seismic waves excited by the excitation source can propagate to most receivers as illustrated in Figure 21. The seismic waves are scattered around the cracks, but those smaller-scale cracks do not hinder the forward propagation of the seismic waves. The subsequent material points can vibrate around their original positions with respect to their neighbor points.

## 5. Conclusions

In this work, seismic response simulation in naturally-fractured media is presented by coupling modified LSM (Liu et al., 2020b) with DFN modeling, a discrete-based numerical approach. Dynamic elastic moduli from seismic responses are calculated and compared with static moduli from uniaxial compression tests. The main conclusions are summarized as follows:

1. The integrated workflow of LSM-DFN is verified to be a surrogate choice to model the seismic responses in complex fractured media by comparing with theoretical predictions. The mechanism of dynamic stress concentration could be further accessed from the aspect of wave scattering theory.

2. Linearly elastic media with a single crack and uniformly oriented cracks show high similarity in the tendency of both the static and dynamic moduli. The equivalent anisotropy obtained by compression tests includes both the inherent and stress-induced parts. Seismic responses can better capture the intrinsic anisotropy of cracked media with optimized receiver arrangement and accurate travel-time estimation.

3. The reason why samples with large fractures tend to be stiffer than samples with homogeneously distributed small fractures can be supplementarily explained by the wavefield theory and visualized by this discrete-based numerical scheme.

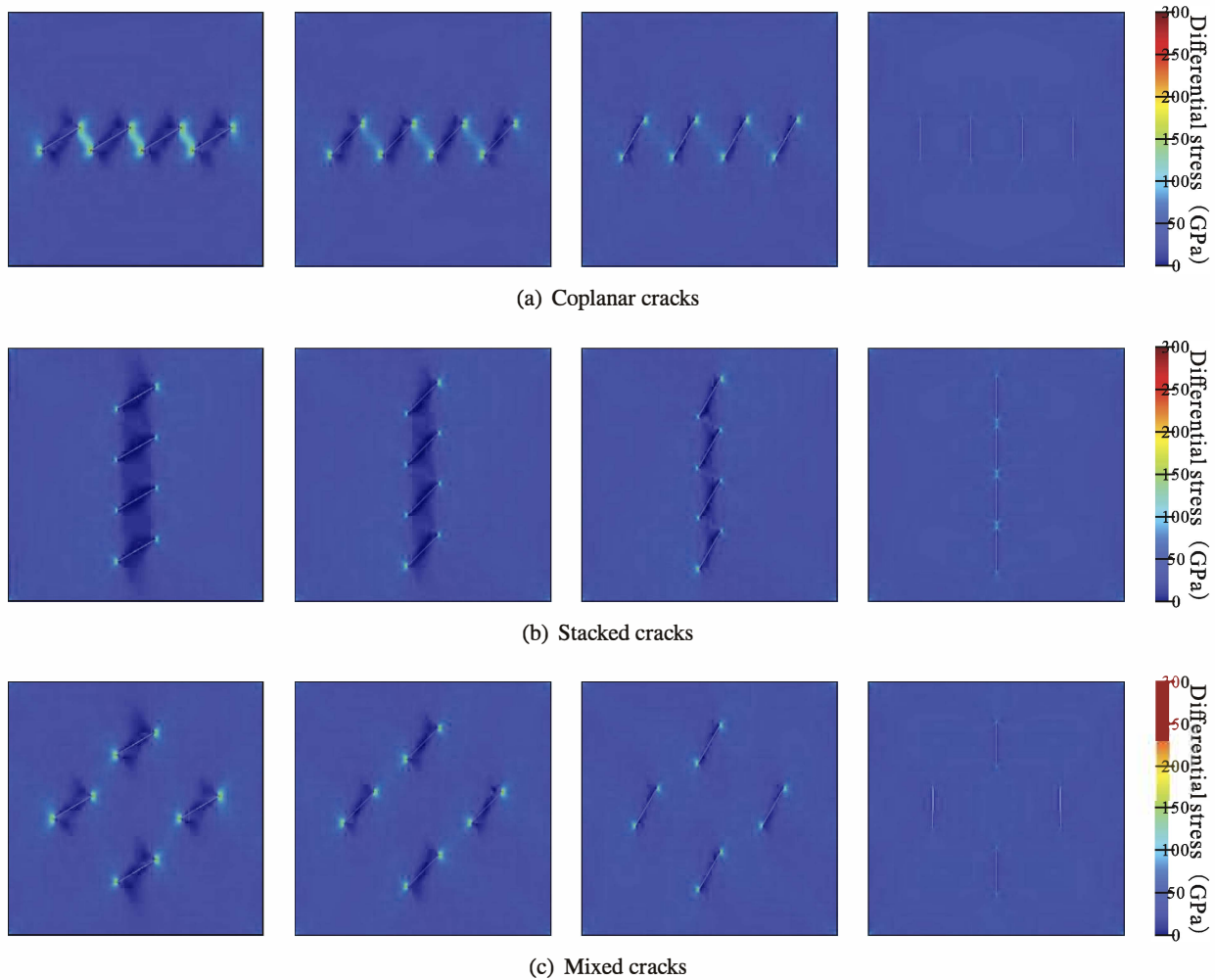

**Figure 14**: Differential stress fields under quasi-static loading state at $\varepsilon_y = 0.60\%$ of LSM-DFN models with a crack half-length of 50 m and four different inclined angles of 30°, 45°, 60°, and 90° in three types of spatial arrangements.

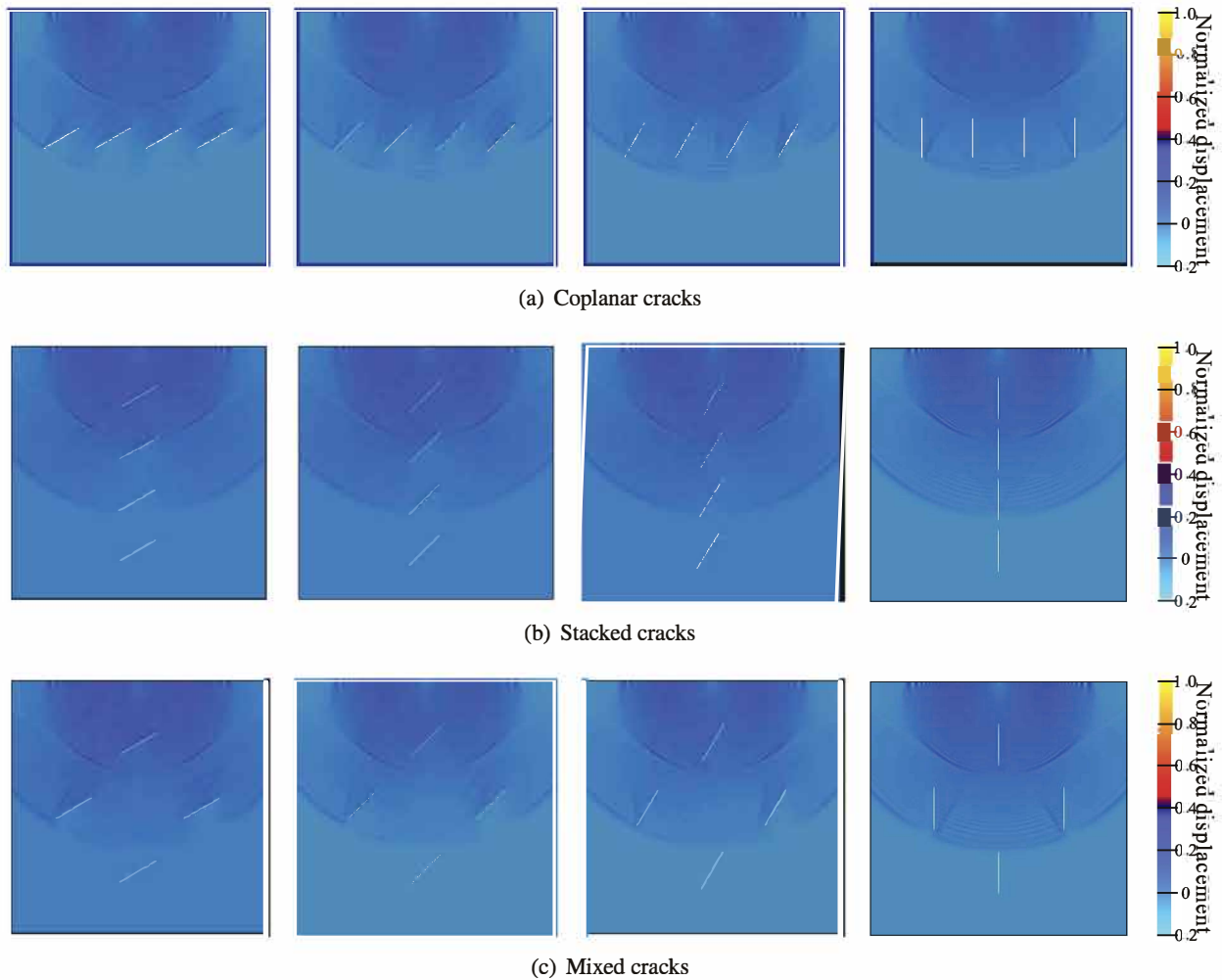

**Figure 15**: Snapshots of the vertical displacement wavefield at $t = 0.70$ s : LSM-DFN models with a crack half-length of 50 m and four different inc ed ngles of 30°, 45°, 60°, and 90° in three types of spatial arrangements.

    

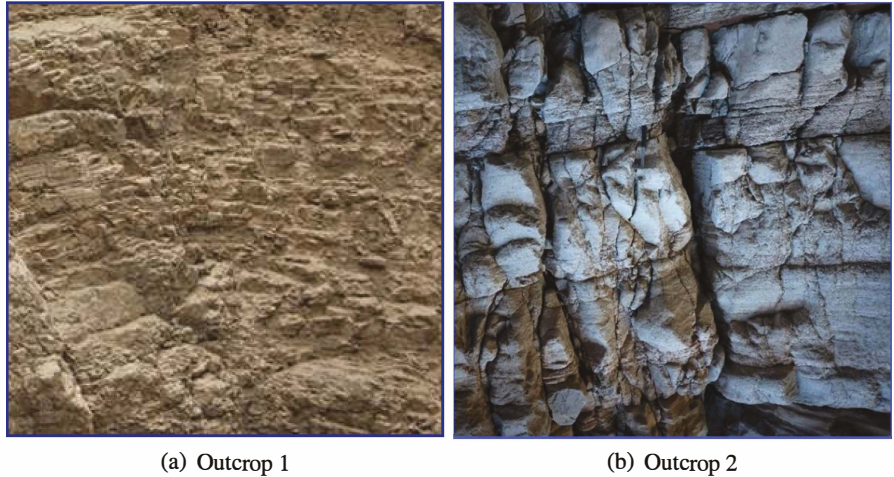

(a) Outcrop 1          (b) Outcrop 2

**Figure 16**: Naturally-fractured media of two realistic outcrops from Ordovician carbonates in the northwest of the Shuntuoguole low uplift of Tarim Basin in China.

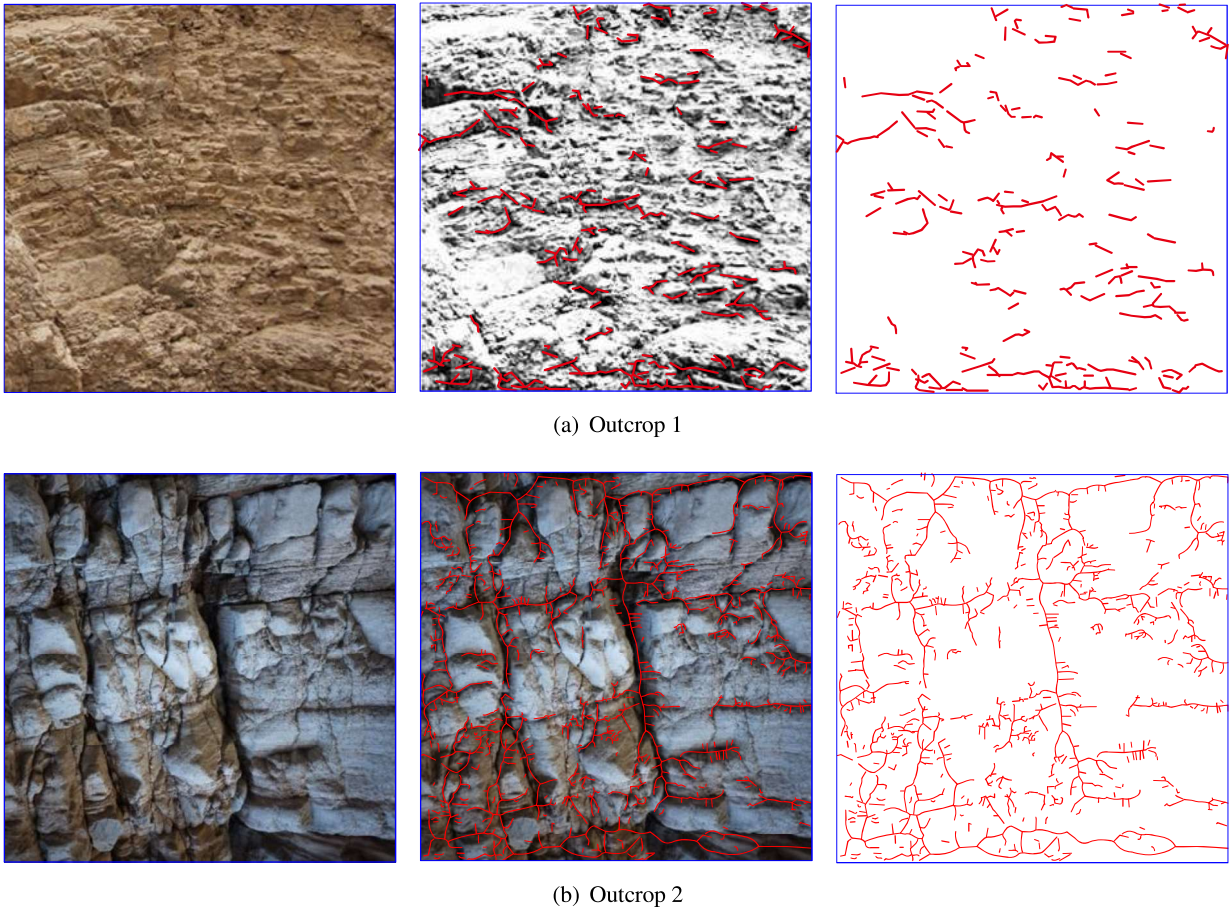

(a) Outcrop 1

(b) Outcrop 2

**Figure 17:** Crack extraction from digital photographs of (a) Outcrop 1 and (b) Outcrop 2 based on GrdHT.

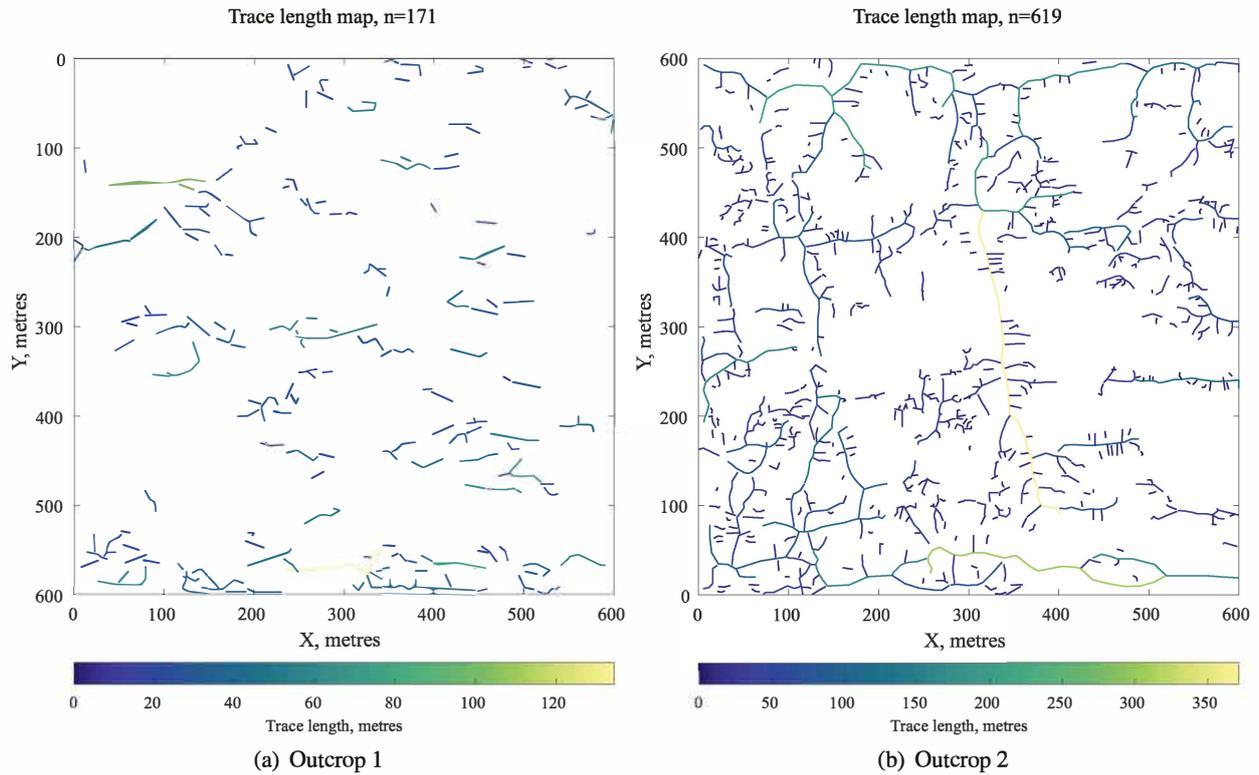

**Figure 18**: Crack length statistics of (a) Outcrop 1 and (b) Outcrop 2.

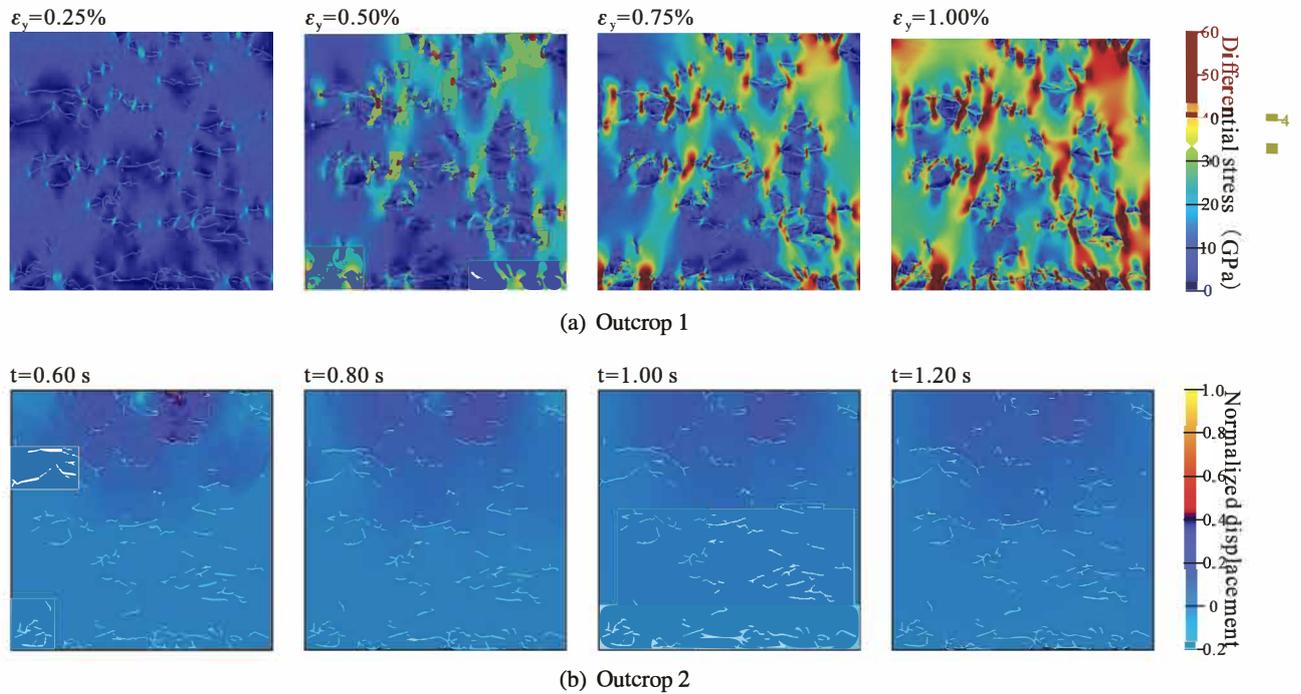

**Figure 19**: Numerical results of Outcrop 1: (a) Differential stress fields under quasi-static loading state at $\varepsilon_y = 0.25\%$, $\varepsilon_y = 0.50\%$, $\varepsilon_y = 0.75\%$, and $\varepsilon_y = 1.00\%$; (b) Snapshots of the vertical displacement wavefield at $t = 0.60$ s, $t = 0.80$ s, $t = 1.00$ s, and $t = 1.20$ s.

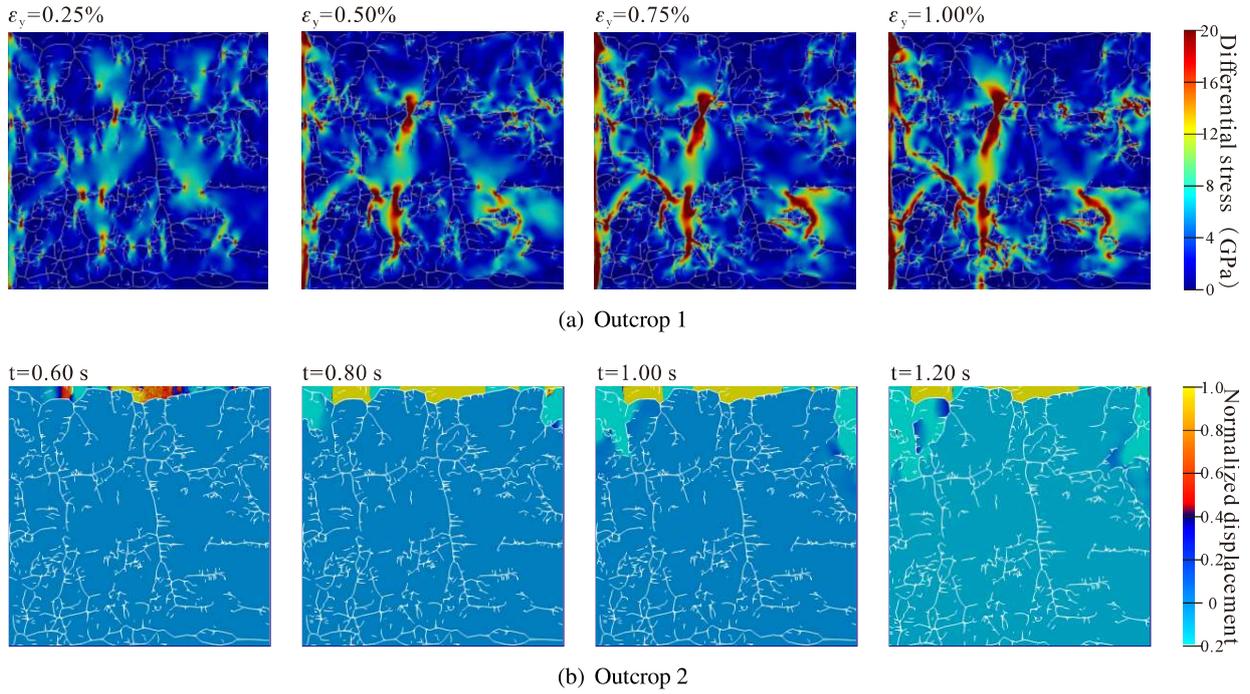

Figure 20: Numerical results of Outcrop 2: (a) Differential stress fields under quasi-static loading state at $\varepsilon_y = 0.25\%$, $\varepsilon_y = 0.50\%$, $\varepsilon_y = 0.75\%$, and $\varepsilon_y = 1.00\%$; (b) Snapshots of the vertical displacement wavefield at $t = 0.60$ s, $t = 0.80$ s, $t = 1.00$ s, and $t = 1.20$ s

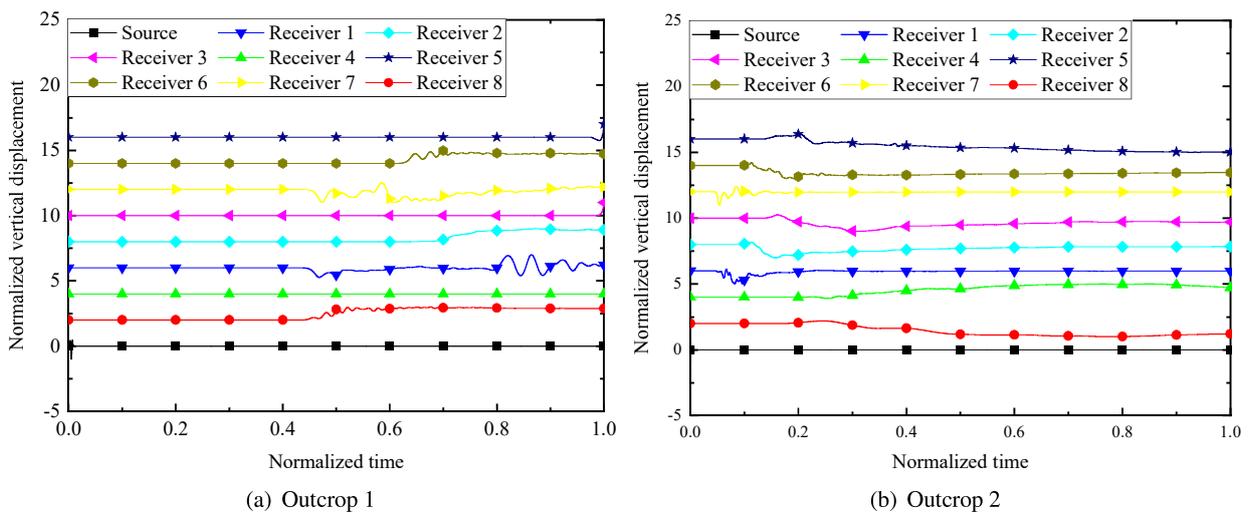

Figure 21: Seismic responses of the source and Receivers 1-7 of (a) Outcrop 1 and (b) Outcrop 2.